\documentclass[conference]{IEEEtran}

\usepackage{cite}
\usepackage{amsmath,amssymb,amsfonts}
\usepackage{graphicx}
\usepackage{textcomp}
\usepackage{xcolor}

\usepackage{booktabs}
\usepackage{enumitem}
\usepackage[most]{tcolorbox}
\usepackage{tikz}
\usetikzlibrary{shapes.geometric, arrows}
\usepackage{subcaption}
\usepackage{array}
\usepackage{url}
\usepackage{hyperref}

\definecolor{lpurple}{RGB}{235,225,255}

\newcommand{\blackcircle}[1]{%
    \tikz[baseline=(char.base)]{
        \node[shape=circle,draw,fill=black,text=white,inner sep=1pt] (char) {#1};
    }%
}

\begin{document}

\title{ARTA: \textbf{A}daptive \textbf{R}einforcement-Learning-Based \textbf{T}hrottling \textbf{A}gent for RowHammer Vulnerabilities}

\author{
\IEEEauthorblockN{Marco Ho}
\IEEEauthorblockA{\textit{British Columbia Institute of Technology} \\
Burnaby, BC, Canada \\
marco\_ho@bcit.ca}
\and
\IEEEauthorblockN{Michael Hsiao}
\IEEEauthorblockA{\textit{Virginia Tech} \\
Blacksburg, VA, United States \\
mhsiao@vt.edu}
\and
\IEEEauthorblockN{Jeeho Ryoo}
\IEEEauthorblockA{\textit{Fairleigh Dickinson University} \\
Vancouver, BC, Canada \\
j.ryoo@fdu.edu}
}

\maketitle

\begin{abstract}
RowHammer vulnerability continues to intensify with DRAM scaling, reducing the activation threshold needed to induce bitflips and rendering existing defenses such as TRR, ECC, and refresh-based mechanisms vulnerable to sophisticated multi-bank hammering patterns. This work presents ARTA, a lightweight reinforcement-learning-based throttling mechanism that detects and suppresses RowHammer activity by monitoring fine-grained memory access behavior within the DRAM refresh window ($t_\text{REFW}$) and dynamically adjusting core throughput using a Q-learning frequency scaling governor. ARTA requires no DRAM-side hardware modification or offline training, using small SRAM structures in the memory controller — a per-core, per-bank FIFO queue (CBF) and a compact Q-table — for immediate deployment. Our evaluation shows that ARTA eliminates all bitflips at $N_\text{BO}$ values down to 64, reduces bitflips up to 22$\text{K}$ times at $N_\text{BO}$ of 20, and improves performance up to 73.6$\%$ over state-of-the-art mitigation mechanisms by limiting preventive action overheads for improved memory bandwidth throughput. These results demonstrate that adaptive RL-based throttling provides robust, scalable, and high-performance RowHammer mitigation for emerging DRAM systems.
\end{abstract}


\section{Introduction}


Since its discovery in 2014~\cite{kim2014flipping}, sustained efforts from the computing community have mitigated RowHammer attacks~\cite{kim2014flipping, kim2020revisiting, kim2024rowhammer, canpolat2024understanding}, yet modern variants bypass even the latest DDR5 systems~\cite{meyer2026phoenix, woo2025qprac, canpolat2025chronus, qureshi2025moat}. RowHammer is a hardware-level read-disturbance mechanism in DRAM that rapidly activates a 
row (i.e., aggressor) to induce bit flips in neighboring rows (i.e., victims) \cite{kim2014flipping}. RowHammer-induced bit flips require a minimum repeated number of row activations in a short timeframe, referred to as the \textit{RowHammer Threshold} ($\text{N}_{\text{RH}}$), which has reduced from 139K in DDR3 \cite{kim2014flipping} to 10K in DDR4 and 4.8K in LPDDR4 \cite{kim2020revisiting}, and is expected to drop below $1\text{K}$ \cite{luo2023rowpress}.
Non-volatile memory (NVM) technologies have also exhibited RowHammer-like disturbance vulnerabilities \cite{agarwal2018spintorque, khan2018sstram, li2014crosspoint, ni2018fefet}. Without effective mitigation techniques, attackers can exploit RowHammer vulnerabilities to compromise data integrity, gain unauthorized access, and escalate privileges. Consequently, research on robust, efficient and scalable RowHammer mitigation strategies is critical to secure sensitive information, preserve user trust, and ensure the reliability of current and future DRAM-based computing systems\cite{kim2014flipping, mutlu2017rowhammer, mutlu2019rowhammer, jattke2022blacksmith, olgun2024abacus, jattke2024zenhammer}.

The computing community has focused on mitigating these attacks by developing various defense strategies \cite{kim2014flipping, park2020graphene, qureshi2022hydra, marazzi2023rega, mutlu2023fundamentally, bostanci2024comet, jattke2024zenhammer, olgun2024abacus, brasser2017canttouchthis, gomez2016dummycells}. In commodity DDR4 chips, Target Row Refresh (TRR) is the most widely adopted in-DRAM RowHammer mitigation. TRR tracks frequently activated aggressor rows and proactively refreshes their neighboring victim rows to prevent bit flips, effectively preventing simple patterns such as double-sided hammering (i.e., two aggressor rows sandwiching one victim row). Unfortunately, recent research~\cite{jattke2022blacksmith, frigo2020trrespass}
has shown that non-uniform hammering patterns can bypass TRR in DDR4 devices, which has led to recently prevailing DDR5 devices integrating more sophisticated mitigation mechanisms such as refresh management (RFM). RFM is a command that provides the DRAM chip with a time window to perform preventive refreshes on potential victim rows. 

As memory systems are subject to naturally occurring faults such as alpha-particle strikes or chip failures, DDR5 DRAM is equipped with On-Die Error Correction Codes (ODECC) to tolerate bit failures \cite{fakhrzadehgan2022safeguard,juffinger2023csi, saxena2023pt, kim2023kill}. However, ODECC does not guarantee correction of all possible failure points, which can lead to data loss. In order to track potential rows targeted by hammering patterns, prior works have proposed RowHammer mitigation mechanisms, each posing relative trade-offs in space and time constraints ~\cite{kim2014flipping, park2020graphene, bennett2021panopticon, qureshi2022hydra, marazzi2023rega, bostanci2024comet, canpolat2024breakhammer, jattke2024zenhammer, olgun2024abacus, qureshi2025moat, canpolat2025chronus, woo2025qprac}. An ideal defense strategy prevents RowHammer attacks by applying clear policies and mechanisms to issue timely preventive refreshes.
These strategies are divided into two types, probabilistic and deterministic, based on the degree of protection they offer against RowHammer attacks, with each approach presenting unique challenges. 

\noindent\textbf{Challenge I: Detection Accuracy-Overhead Tradeoff.}
Probabilistic trackers~\cite{kim2014flipping,saxena2022aqua,jaleel2024pride,qureshi2024mint} approximate victim rows with minimal storage, but their inherent randomness issues redundant refreshes of non-critical rows. PrIDE~\cite{jaleel2024pride} and MINT~\cite{qureshi2024mint} each incur ${\approx}30\%$ activation bandwidth loss at $N_\text{RH}=250$~\cite{woo2025qprac}, making them impractical at sub-1K thresholds.

\noindent\textbf{Challenge II: Scalability of Row Tracking.}
Counter-based deterministic solutions~\cite{olgun2024abacus,park2020graphene,bennett2021panopticon,qureshi2022hydra,bostanci2024comet,taneja2025dream} provide precise tracking but face storage and performance bottlenecks as $N_\text{RH}$ falls. Shared-counter schemes~\cite{park2020graphene,qureshi2022hydra} introduce blind spots exploitable by multi-bank attacks, while Panopticon's FIFO-based tracking~\cite{bennett2021panopticon} cannot prioritize the most critical rows~\cite{woo2025qprac,qureshi2025moat}.

\noindent\textbf{Challenge III: Software Mitigation Limitations.}
Software defenses~\cite{zhang2022softtrr,van2018guardion,konoth2018zebram,aweke2016anvil,bock2019rip} avoid hardware modifications but lack the fine-grained control needed to track hardware-level attacks and require proprietary DRAM parameters unavailable from vendors. Recent attacks~\cite{gruss2016rowhammer,cojocar2019exploiting} have bypassed most such solutions.

\noindent\textbf{Toward Adaptive RowHammer Defense.}
As attacks grow more sophisticated~\cite{fiedler2026memory}, static and fixed-threshold defenses cannot keep pace. Reinforcement learning (RL) offers a promising direction: RL agents learn optimal policies through environment interaction, providing continual adaptability without offline training~\cite{biswas2024ai,panda2024reinforcement,ojha2025adaptive}. To the best of our knowledge, no RL-based approach has been proposed for RowHammer mitigation.


\begin{tcolorbox}[colback=lpurple,
                  colframe=black,
                  breakable,
                  enhanced,
                  sharp corners]
\textbf{Goal.} To design a scalable, adaptive, and deployable RowHammer defense using a lightweight RL-based algorithm to detect and mitigate emerging hammering patterns, eliminating the need for hardware changes or offline training.

\textbf{Key Idea.} To dynamically throttle the frequency of memory requests from a hardware thread by approximating repeated and close-proximity memory accesses.

\textbf{Key Mechanism.} ARTA, a first-of-its-kind lightweight RL-based Dynamic Frequency Scaling (DFS) governor that utilizes Q-learning to detect and mitigate RowHammer in real time through pattern-aware throttling. Detection is modeled as a Markov Decision Process (MDP)~\cite{rao2000reinforcement}, and a pre-initialized Q-table with Gaussian-like decay removes exploration overhead.
\end{tcolorbox}





Source-level DFS throttles attack-generated memory traffic at its origin by reducing CPU clock frequency, cutting row activations before $N_\text{BO}$ is reached. We replace the conventional ACPI DFS governor~\cite{sciencedirect_acpi_overview,hebbar2022pmu} with an RL-based controller that detects RowHammer patterns via second-order differences of per-bank FIFO address queues and selects optimal P-states or the C1 idle state in response. This paper makes the following contributions:

\begin{enumerate}[leftmargin=12pt]
    \item \noindent\textbf{Dynamic Pattern Detection}: We show repeated, close-proximity memory accesses can be accurately estimated to distinguish RowHammer patterns from benign workloads.
    \item \noindent\textbf{Adaptive Throttling}: We present the first RL-based RowHammer throttling mechanism that adapts to real-time hammering patterns over changing workload demands. 
    \item \noindent\textbf{Immediate Deployment}: We pre-initialize the Q-table to accelerate policy convergence, achieving $100\%$ precision.
    \item \noindent\textbf{Reduced System Overheads}: We evaluate ARTA, reducing bitflips by $22\text{K}\times$ with up to $73.6\%$ speedup.
    \item \noindent\textbf{Minimal Chip Area}: We present a lightweight, state-of-the-art RowHammer defense that preserves low area overhead ($1.4$KB/core) with decreasing $N_\text{RH}$ as low as $20$.
\end{enumerate}

\section{Background}

\label{sec:background}
\subsection{DRAM Organization and Operations}

DRAM memory systems are organized hierarchically as a set of ranks, banks, and cells. 
The operation of DRAM is initiated by a memory controller in the processor. Memory channel buses transport incoming DRAM memory requests, each defined with specific timing constraints that the memory controller uses to coordinate data transfer with a rank in DRAM. Each rank contains independent DRAM banks operating in lockstep. A part of memory address bits is used to index the rank, the bank in that rank, and the memory array of cells organized by rows and columns in each bank.


To access data stored in a DRAM cell, the memory controller issues an activation (ACT) command, specifying both the bank address and row address associated with the data address. The data is latched in the row buffer where data is read by supplying a column address. To access another row (i.e., \textit{row conflict}), the memory controller must issue a precharge (PRE) command to close the currently open row first. The row must remain open for at least tRAS time before it can be closed for subsequent access to another row.


To maintain data integrity, periodic refresh (REF) operations are required to restore the charge for each cell capacitor. These REF operations must occur within a refresh window ($t_{\text{REFW}}$). To systematically refresh each row at least once within the $t_{\text{REFW}}$ window, the memory controller issues a REF command for one row of DRAM cells in all banks at regular refresh intervals ($t_\text{REFI}$) \cite{jacob2010memory}, typically every $7.8 \,\mu\text{s}$ for a 64\,ms window, or $3.9 \,\mu\text{s}$ for a 32\,ms window \cite{jedecddr4, jedecddr3, micronddr4, jedeclpddr4, jedecddr5, jedeclpddr5}. To refresh all cells within $t_{\text{REFW}}$, the memory controller cycles through every row to issue REF commands in a round-robin or interleaved manner, preventing any single row from exceeding the maximum allowable time without a refresh. When a row is refreshed, there is a refresh latency ($t_{\text{RFC}}$) that temporarily pauses normal memory access operations in that row to complete the refresh operation.

\subsection{RowHammer Mechanism}
\label{subsec:read-disturb}
RowHammer is a security vulnerability in which rapid repeated activations to an aggressor row induce read disturbance bitflips in nearby victim rows due to electrical interference \cite{kim2014flipping, mutlu2017rowhammer, mutlu2019rowhammer, kim2020revisiting, mutlu2023fundamentally, olgun2025variable,kang2024sledgehammer}. This phenomenon, known as read disturbance, is caused by capacitive coupling. Frequently opening (ACT) and closing (PRE) the aggressor row causes charge leakage in storage capacitors of adjacent ($\pm1$) or even non-adjacent ($\geq\pm2$) rows up to a blast radius (BR) of $\pm2$ to $\pm4$ rows. If a row is activated at least $N_{\text{RH}}$ times within $t_\text{REFI}$, data in nearby rows may be corrupted before they are refreshed. With technology scaling, prior works show $N_{\text{RH}}$ has dropped from 139K in DDR3 \cite{kim2014flipping} to 10K in DDR4 and 4.8K in LPDDR4 \cite{kim2020revisiting}, and is expected to decline below $1\text{K}$ \cite{luo2023rowpress}.


\subsection{Mitigation Approaches}
\label{subsec:mit-app}
In DDR4 and DDR5, TRR has been widely adopted as an in-DRAM RowHammer mitigation. TRR tracks frequently activated aggressor rows and proactively refreshes their neighboring victim rows to prevent bit flips. Prior works have proposed read disturbance mitigation mechanisms to prevent bitflips from RowHammer attacks~\cite{kim2014flipping, park2020graphene, bennett2021panopticon, qureshi2022hydra, marazzi2023rega, bostanci2024comet, canpolat2024breakhammer, jattke2024zenhammer, olgun2024abacus, qureshi2025moat, canpolat2025chronus, woo2025qprac}. Consequently, memory controller-based solutions complement TRR by extending the tracking and mitigation capabilities beyond what is possible within DRAM chips alone~\cite{kim2024rowhammer}. These solutions rely on probabilistic or deterministic triggers, each offering different trade-offs in terms of storage and timing overheads.

\noindent\textbf{Probabilistic Mitigation.} Probabilistic mechanisms often achieve lower chip area overhead by using approximation techniques such as random sampling or probabilistic row refreshes~\cite{kim2014flipping, jaleel2024pride, qureshi2024mint, saxena2022aqua}. However, their inherent randomness leads to imperfect coverage, resulting in redundant refreshes of non-critical rows. As a result, such approaches do not scale well for lower $N_{\text{RH}}$ thresholds, where more precise tracking becomes critical.

\noindent\textbf{Deterministic Mitigation.} In contrast, deterministic solutions maintain accurate access tracking by leveraging activation counters, usually incurring higher storage overhead~\cite{kim2024rowhammer}. To reduce this overhead, many works propose shared counters across groups of rows~\cite{park2020graphene, qureshi2022hydra, bostanci2024comet, olgun2024abacus}. However, shared counters struggle to deliver fine-grained tracking, especially as DRAM densities rise and thresholds fall. For instance, in Hydra~\cite{qureshi2022hydra}, shared counters are synchronized periodically with individual counters, introducing computational overhead and delayed refresh accuracy. Similar issues arise in DREAM~\cite{taneja2025dream}, ABACuS~\cite{olgun2024abacus}, CoMeT~\cite{bostanci2024comet}, and Graphene~\cite{park2020graphene}, where tracking is approximate and vulnerable to adversarial patterns such as multi-bank attacks~\cite{kang2024sledgehammer}. These patterns can overload shared counters, forcing redundant all-bank refreshes, reducing DRAM bandwidth and increasing slowdown~\cite{woo2025dapper, olgun2024abacus}.



\begin{figure}[t]
    \centering
    \includegraphics[width=1\linewidth]{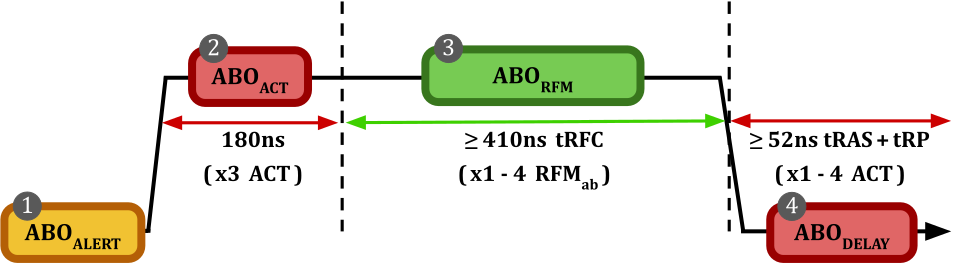}
    \caption{Overview of ABO timing in PRAC.}
    \label{fig:abo-timing}
    \vspace{-6mm}
\end{figure}

\noindent\textbf{DDR5 PRAC+ABO.} To standardize and support accurate row-level tracking in DRAM, the April 2024 JEDEC update of the DDR5 specification introduced PRAC \cite{jedecddr5c}, a new on-DRAM-die read-disturbance error (RDE) mechanism that maintains an ACT counter for each row to precisely track activations. Illustrated in Fig. \ref{fig:abo-timing}, when a row's ACT count reaches a back-off threshold ($N_\text{BO}$) \footnote{$N_\text{BO}$ can be configured to 70\%, 80\%, 90\%, or 100\% of $N_{\text{RH}}$ \cite{jedecddr5c}}, \blackcircle{1} the DRAM chip asserts an alert back-off (ABO) signal to the MC, \blackcircle{2} permitting additional activations ($\text{ABO}_{\text{ACT}}$) within a limited window (e.g., $180~\text{ns}$), before \blackcircle{3} performing a fixed number of RFM commands ($\text{ABO}_{\text{RFM}}$), and \blackcircle{4} enforcing a minimum activation quota prior to the next ABO event ($\text{ABO}_{\text{DELAY}}$). Table \ref{tab:prac-params} summarizes PRAC-related parameters.

Building on TRR, RFM is a DDR5-specified DRAM command that allocates a time window (e.g., $t_{\text{RFM}} = 410~\text{ns}$) for the DRAM chip to perform preventive refreshes of potential victim rows. Specifications prior to 2024 (e.g., DDR5 \cite{jedecddr5}) recommend that the memory controller trigger RFM when total row activations in a bank or logical memory region exceed a threshold (e.g., $N_\text{BO} = 32$ \cite{jedecddr5c}).



\subsection{Variants of RowHammer}
\label{subsec:rh-variants}
Recent RowHammer attacks on DDR4 and DDR5 have demonstrated that TRR is exploitable using crafted memory access patterns that bypass existing mitigation techniques \cite{kim2014flipping, mutlu2017rowhammer, mutlu2019rowhammer, kim2020revisiting, mutlu2023fundamentally, olgun2025variable}, including single-sided, double-sided, multi-sided, and multi-bank hammering. Sophisticated multi-sided attacks such as Half-Double \cite{kogler2022half} and BLASTER \cite{lang2023blaster} have been shown to induce bitflips in non-adjacent victim rows (i.e., $\pm2$ - $4$ BR). Other sophisticated attacks such as TRRespass \cite{frigo2020trrespass} and BlackSmith\cite{jattke2022blacksmith} exploit the timing constraints of refresh-based mitigations using complex, non-uniform memory access patterns to induce bitflips.

\begin{figure}[h]
\centering
\includegraphics[width=1.0\linewidth]{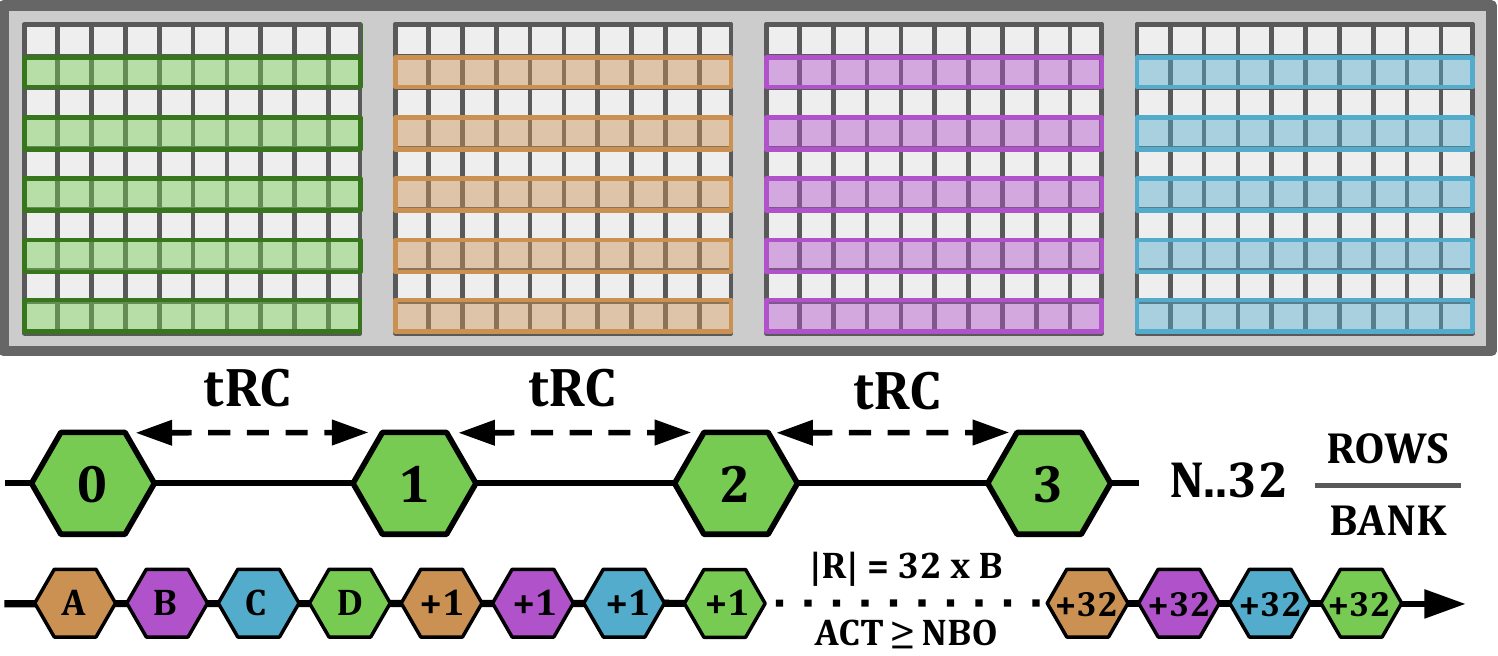}
\caption{Overview of 32-sided Multi-Bank RowHammer Attack.}
\label{fig:rh32-attack}
\vspace{-3mm}
\end{figure}

Recent multi-bank attacks such as SledgeHammer \cite{kang2024sledgehammer} further accelerate bit flips on DDR4 by leveraging bank-level parallelism and cache self-eviction techniques. 
Fig.~\ref{fig:rh32-attack} illustrates a 32-sided, multi-bank RowHammer pattern that interleaves aggressor activations across all banks. During the setup phase, each bank is sequentially hammered in 32 consecutive aggressor rows, with activations interleaved so that each bank receives its next activation in turn. Every aggressor row is driven up to $N_{\text{BO}} - 1$ activations (ATH). During the attack phase, under PRAC, an ABO event is induced in an ATH row per round, and the resulting $\text{ABO}_{\text{ACT}}$ and $\text{ABO}_{\text{DELAY}}$ activations distribute across the set of remaining ATH row in multiple rounds to ultimately exceed $N_{\text{RH}}$, inducing bitflips.


\begin{table}[t]
    \centering
    \caption{PRAC Parameters as per DDR5 Specification \cite{jedecddr5c}}
    \label{tab:prac-params}
    \scriptsize
    \begin{tabular}{
        @{}
        >{\raggedright\arraybackslash}m{0.11\linewidth}
        >{\centering\arraybackslash}m{0.41\linewidth}
        >{\centering\arraybackslash}m{0.38\linewidth}
        @{}
    }
        \toprule
        \textbf{Parameter} & \textbf{Explanation} & \textbf{Value} \\
        \midrule
        $N_\text{BO}$ & Back-Off Threshold & $N_\text{BO} \leq N_{\text{RH}}$ \\
        $\text{ABO}_{\text{ACT}}$ & Max. ACTs from Alert to RFM & 3 (up to 180ns) \\
        $\text{ABO}_{\text{RFM}}$ & Max. RFMs on Alert & 1, 2, or 4 \\
        $\text{ABO}_{\text{DELAY}}$ & Min. ACTs after RFM to Alert & Same as $\text{ABO}_{\text{RFM}}$ (1,2, or 4) \\
        $t_{\text{RFM}}$ & Duration of $\text{RFM}_\text{ab}$ & 410ns \\
        \bottomrule
    \end{tabular}
    \vspace{-3mm}
\end{table}

\vspace{-2mm}

\subsection{Reinforcement Learning Fundamentals}
\label{subsec:rl-solutions}

\begin{figure}[t]
\vspace{0mm}
\centering
\includegraphics[width=1.0\linewidth]{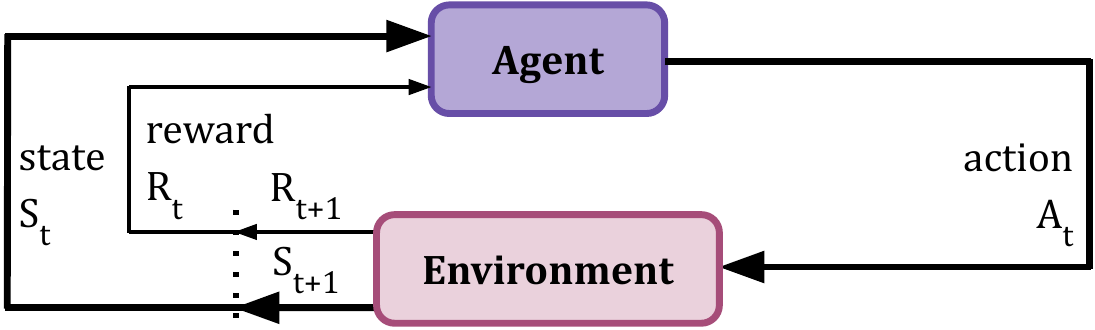}
\caption{Overview of MDP in Reinforcement Learning.}
\label{fig:rl-logic}
\vspace{-7mm}
\end{figure}

Fundamentally, RL is the algorithmic approach of learning how to select an action in a given situation to maximize a numerical reward signal. A typical RL system is modeled as a Markov Decision Process (MDP) comprising an agent and an environment. The agent is the entity that selects actions by interacting with an environment in discrete timesteps, as shown in Fig. \ref{fig:rl-logic}. At each timestep $t$, the agent (1) chooses an \textbf{\textit{action}} $a_{t}$ based on the current \textbf{\textit{state}} $s_{t}$, causing the environment to transition to a new state $s_{t+1}$ and (2) emit an immediate or delayed \textbf{\textit{reward}} ($r_{t+1}$) at the next state transition.

The agent's objective is to maximize cumulative reward via an optimal policy. The expected cumulative reward for each state-action pair is the \textbf{\textit{Q-value}} $Q(s,a)$, stored in a Q-table. Q-values are updated via the \textit{Bellman Optimality Equation}:

{\footnotesize
\[
\begin{aligned}
Q(s_t, a_t) \leftarrow\; & Q(s_t, a_t) \\
& \hspace{-1em} + \alpha \left[ r_{t+1} + \gamma \max_{a'} Q(s_{t+1}, a') - Q(s_t, a_t) \right]
\end{aligned}
\]
}

\noindent Here, $\alpha$ is the \textit{learning rate} (convergence speed) and $\gamma$ the \textit{discount factor} (balance between immediate and future rewards).

To avoid suboptimal greedy policies~\cite{rao2000reinforcement}, the $\epsilon$-greedy strategy selects a random action with probability $\epsilon$ (the \textit{exploration rate}), which decays over time to shift from exploration toward exploitation.

\section{Motivation}

As the physical distance between adjacent DRAM rows shrinks with increasing chip density, susceptibility to RowHammer and related read disturbance errors (RDEs) increases due to higher cell-to-cell coupling and accelerated charge leakage. Mitigating these vulnerabilities requires more aggressive refresh rates, which impose prohibitive performance overheads at decreasing RowHammer thresholds. These refresh-based approaches assume the memory controller can react rapidly enough to prevent critical charge loss. However, as thresholds are expected to decline below $1\text{K}$ activations~\cite{luo2023rowpress}, reliable data retention cannot be guaranteed by current timing guardbands in DRAM, motivating the need for novel detection and mitigation strategies that address these fundamental scaling challenges.

These scaling challenges are compounded by the continued evolution of RowHammer attack patterns. Early DDR4 mitigation mechanisms used coarse-grained tracking with shared counters to identify vulnerable rows \cite{park2020graphene,qureshi2022hydra,bostanci2024comet,olgun2024abacus}, which minimized area overheads but introduced blind spots. As attack techniques progressed beyond single- and double-sided hammering to multi-sided and multi-bank patterns, these blind spots became increasingly exploitable. To address these limitations, the latest DDR5 standard introduced PRAC, which provides fine-grained tracking with on-die per-row activation counters in each bank. When a counter exceeds $N_\text{BO}$, PRAC issues an ABO signal that triggers RFM (\S\ref{subsec:mit-app}).


However, modern RowHammer attacks are increasingly dynamic and sophisticated, continually bypassing heuristic-based mitigation strategies designed to detect known patterns. This adaptability renders static, fixed-threshold defenses, such as current DDR5 refresh mechanisms, insufficient to handle increasingly complex, multi-bank, and dynamic RowHammer attacks \cite{meyer2026phoenix}. As physical scaling exacerbates RowHammer susceptibility, and empirical evidence shows PRAC’s limitations at lowered $N_\text{BO}$ thresholds \cite{canpolat2024understanding, woo2025qprac, canpolat2025chronus, qureshi2025moat}, it is clear prior approaches do not fully address the evolving attack landscape.



\begin{figure}[t]
    \centering
    \includegraphics[width=1\linewidth]{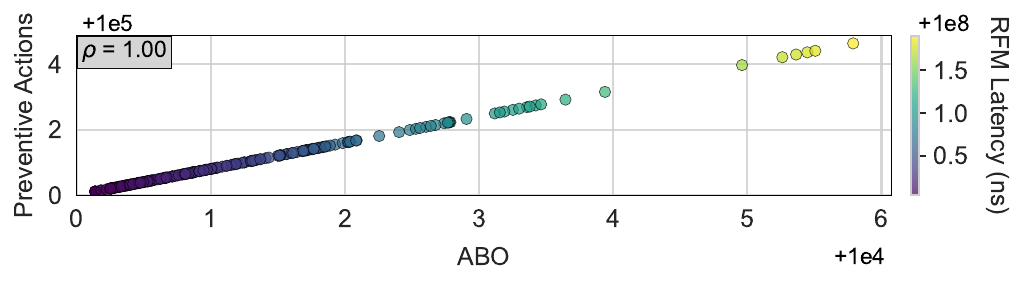}
    \caption{PRAC's effect on ABOs, preventive actions, and RFM latency in $56$ multi-bank attack workloads ($N_\text{BO}$ of $128$ to $20$).}
    \label{fig:bivariate}
    \vspace{-5mm}
\end{figure}

Unfortunately, as $N_\text{BO}$ reduces, PRAC becomes increasingly vulnerable to channel-wide lockouts at sub-$1\text{K}$ activation thresholds~\cite{kim2020revisiting}. Between ABOs, PRAC enforces fixed activations from $\text{ABO}_{\text{ACT}}$ and $\text{ABO}_{\text{DELAY}}$ on each refreshed bank, which increment row counters that may trigger subsequent ABOs if the counters reach $N_\text{BO}$. The ABO signals lead to issuing all-bank RFM commands, with Fig. \ref{fig:bivariate} illustrating a near linear correlation of ABO and preventive actions (RFM) on cumulative RFM latency.
Performing RFM on each bank from ABO increases ABO frequency because PRAC issues fixed activations across refreshed banks on every ABO, which raises the likelihood of RDE bitflips. Given these limitations, an alternative approach is needed to better address the vulnerabilities introduced by reduced $N_\text{BO}$. One such approach is Chronus~\cite{canpolat2025chronus}, 
a PRAC-based mechanism that eliminates $\text{ABO}_{\text{ACT}}$ and $\text{ABO}_{\text{DELAY}}$ constraints. 

\begin{figure}[h]
    \centering
    \includegraphics[width=1\linewidth]{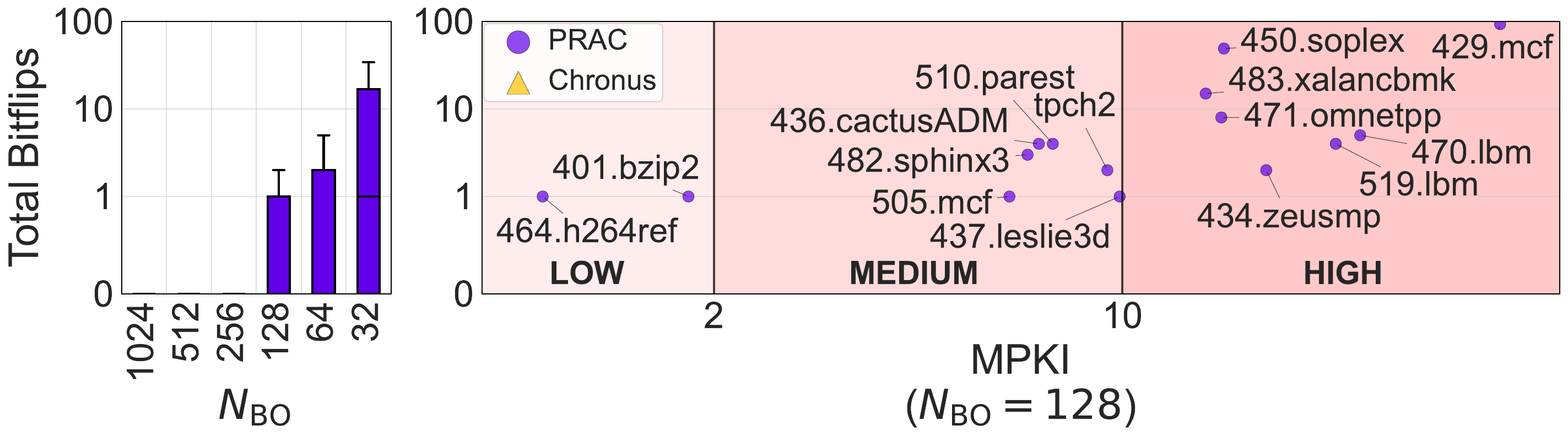}
    \caption{Bitflips from $56$ benign workloads in (a) PRAC across $N_\text{BO}$ (left) and (b) in PRAC and Chronus at $N_\text{BO}=128$ (right).
    }
    \label{fig:bitfips-motivation}
    \vspace{0mm}
\end{figure}

We evaluate the security implications of PRAC and Chronus as $N_\text{BO}$ decreases from $1024$ to $128$. Fig. \ref{fig:bitfips-motivation} illustrates the growing number of bitflips in PRAC across decreasing $N_\text{BO}$ thresholds, confirming that refresh alone cannot fully protect benign workloads from RDE vulnerabilities. These results are presented as a box-and-whisker plot \footnote{The box extends from the first quartile (the median of the lower half of the ordered data) to the third quartile (the median of the upper half). The interquartile range ($IQR$), representing the box height, is the difference between these quartiles. Whiskers extend to the most extreme data points within $1.5 \times IQR$ from the quartiles.} across evaluated $N_\text{BO}$ values (a) and as a scatterplot of bitflips at $N_\text{BO}$ of 128 (b).
We make two observations from Fig. \ref{fig:bitfips-motivation}a. First, bitflips begin manifesting at an $N_\text{BO}$ of 128, for a total of 193 across all workloads. Second, we observe increased RDE susceptibility in PRAC as $N_\text{BO}$ decreases, whereas no bitflips were observed in Chronus. As shown in Fig. \ref{fig:bitfips-motivation}b, PRAC exhibited bitflips in $15$ of $56$ workloads at $N_\text{BO}$ of 128, reaching up to $93$ bitflips (\textit{429.mcf}). These results indicate that PRAC’s ability to mitigate bitflips diminishes as $N_\text{BO}$ decreases, further emphasizing the limitations of refresh-based mitigation techniques. Although fixed or hard-coded mitigation policies such as PRAC provide baseline protection, they lack the flexibility to adapt to dynamic workload behaviors and attack patterns.

Taken together, these observations reveal that existing deterministic refresh mechanisms and fixed mitigation policies cannot keep pace with highly sophisticated RowHammer attacks, particularly at low $N_\text{BO}$ thresholds. The increasing vulnerability to multi-bank attack patterns further complicates the challenge by expanding the attack surface, underscoring the potential for increasingly dynamic and complex RowHammer attacks in the future. A mechanism that can adaptively respond to changing attack strategies and workload behavior is therefore essential. To close this gap, we propose ARTA, an adaptive reinforcement-learning-based throttling agent that dynamically adapts to changing workload behavior and attack patterns to mitigate RowHammer attacks more effectively than deterministic mitigation policies.





\section{ARTA Implementation Details}



Our RL-guided DFS framework, ARTA, adaptively throttles cores exhibiting RowHammer-like access patterns by replacing the conventional frequency governor with an RL-driven OS-level controller. Processor state transitions are aligned with the DDR5 refresh window ($t_{\text{REFW}}=32$ ms) to mask DFS latency. At each window, ARTA observes memory activity, quantifies its severity, selects a P- or C-state action, and receives a delayed reward reflecting the system response. States represent discretized levels of RowHammer-relevant access intensity, derived from per-core address queues. Q-learning updates a compact Q-table to map observed patterns to optimal throttling decisions, enabling real-time attack suppression with minimal performance and energy overhead.


\subsection{ARTA Architecture}

\begin{figure}[htbp]
    \vspace{-2mm}
    \centering
    \includegraphics[width=1\linewidth]{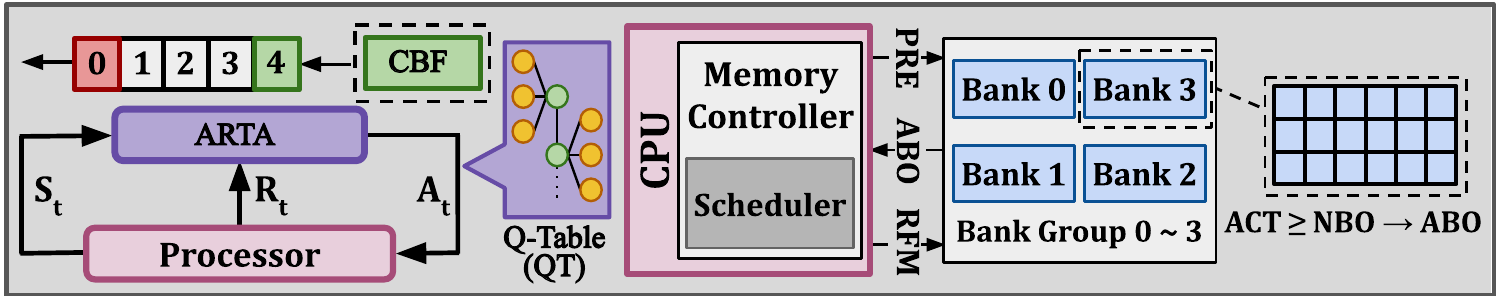}
    \caption{Overview of DRAM architecture with ARTA. ARTA adds two hardware structures to the MC: the \textbf{CBF} (per-core, per-bank FIFO, 16 entries $\times$ 17-bit row IDs; shown as \textit{Row X}) and the \textbf{Q-Table (QT)} (11 states $\times$ 6 actions; shown as the RL network). PRE requests are indexed by core ID and bank ID to update the CBF, which computes the severity score fed to the QT.}
    \label{fig:arta-arch}
    \vspace{-3mm}
\end{figure}

Fig. \ref{fig:arta-arch} shows a high-level overview of ARTA. ARTA consists of two key hardware structures: Q-Table (QT) and per-core, per-bank FIFO queues (CBF). The QT stores Q-values $Q(s,a)$ for each state-action pair $(s,a)$. Each CBF maintains a list of per-bank first-in-first-out (FIFO) queues and registers that store three key pieces of information: (1) the current state of the core, (2) the action taken, and (3) the clock cycle when throttling occurred. The recorded clock cycle ensures that throttling adheres to the minimum delay required by the ACPI specification, thus preventing the CPU from being throttled earlier than permitted.

The CBF tracks recent memory addresses from PRE requests. For each PRE request, the requesting CPU core ID and the corresponding DRAM bank ID are parsed from the request metadata to efficiently index the request to the appropriate CBF. By organizing each request based on the requesting core ID and its associated bank ID, ARTA disentangles inter-core and inter-bank contention, enabling accurate computation of second-order differences for detection of hammering patterns.

ARTA employs second-order difference computation of recently accessed memory addresses to detect RowHammer attack patterns. A RowHammer attack pattern exhibits two inherent characteristics: (1) repetition, where aggressor rows are repeatedly activated within a short period of a single DRAM refresh window ($t_{\text{REFW}}$) of $32\,\text{ms}$ (DDR5 \cite{jedecddr5c}) or $64\,\text{ms}$ (DDR4 \cite{jedecddr4}), and (2) proximity, where consecutive memory accesses target rows physically adjacent to victim rows. By using the CBF to continuously monitor recent memory access sequences, ARTA captures these temporal and spatial correlations to distinguish hammering patterns from benign patterns.

\subsection{RL Control Mechanisms}
\subsubsection{\textbf{Q-table Initialization}}
The discrete states ($s \in [0,1]$ with cardinality $N_S$) represent the severity of the pattern (e.g., 0\%, 10\%, 20\%), quantifying repeated and close-proximity memory accesses; values closer to 1 indicate higher severity. Actions ($a \in [0,1]$ with cardinality $N_A$) correspond to discrete processor states, that is, P-states or, for worst-case patterns, an idle C-state (C1) to adjust CPU core clock frequency with DFS in response to changing access patterns.

\begin{figure}[t]
    \centering
    \includegraphics[width=1\linewidth]{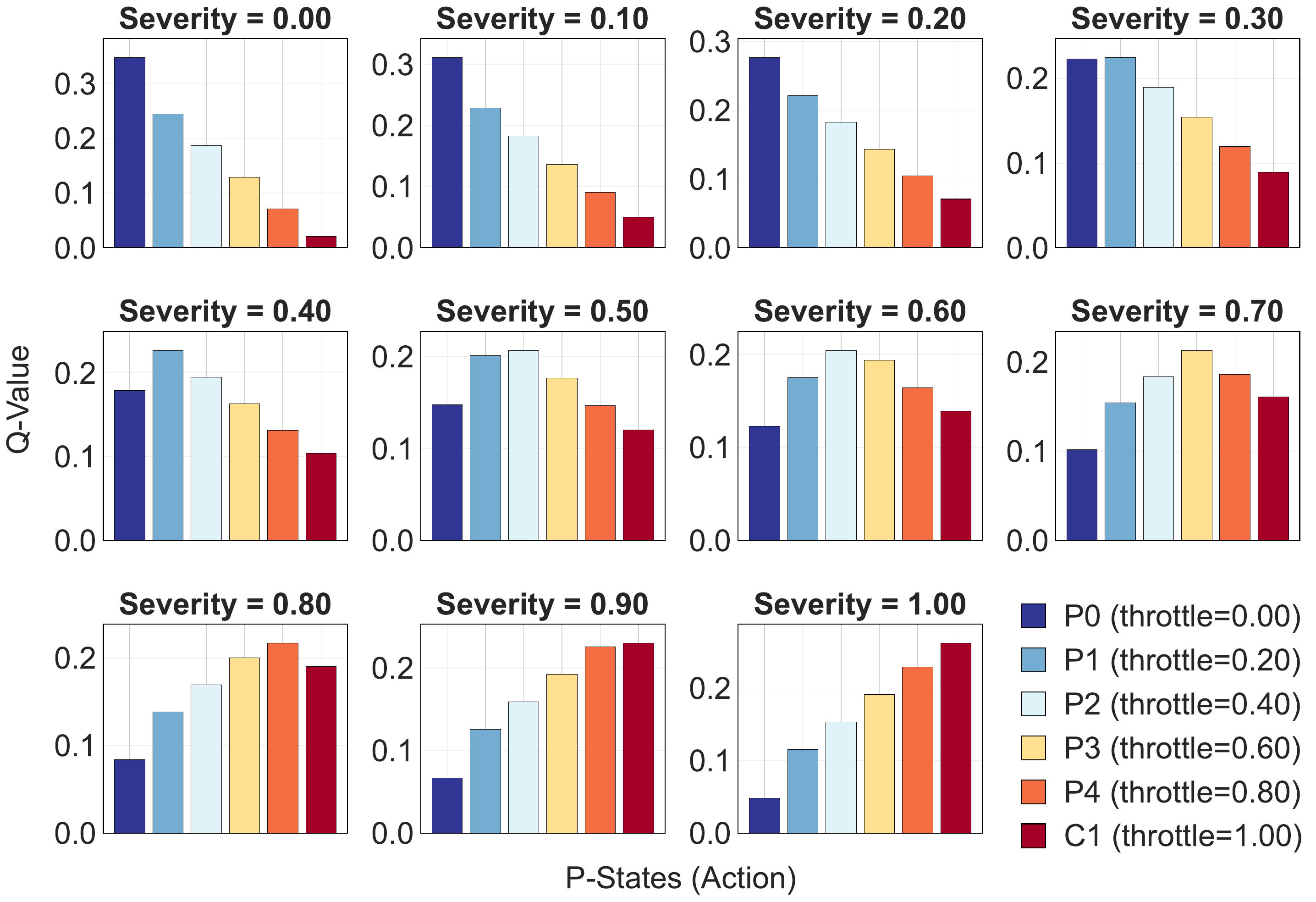}
    \caption{Gaussian-like linear-decay for pre-initializing Q-values.}
    \label{fig:q_values}
    \vspace{-4mm}
\end{figure}

To accelerate convergence, the Q-value of each state-action pair in the QT is pre-initialized using a Gaussian-like linear decay function. Pre-initialization reduces the exploration time and accelerates convergence by mapping actions such as P0 to discretized states (e.g., 0\%, 10\%, 20\%) as shown in Fig. \ref{fig:q_values}.

Given each action $a$, we define a preferred severity interval $[s_a^{\min}, s_a^{\max}]$. The first (P0) and last (C1) actions are assigned boundary intervals to cover the extremal state values, where $s_0^{\min} = 0$, $s_0^{\max} = s_{\min}$ for P0, and $s_{\text{1}}^{\min} = s_{\max}$, $s_{\text{1}}^{\max} = 1$ for C1. The remaining interior actions (P1, P2, P3, P4) are assigned equally spaced intervals over $[s_{\min}, s_{\max}]$. The center of each interval, $\bar{s}_a = (s_a^{\min} + s_a^{\max})/2$, represents the target severity of action $a$.

For a given state $s$, the initial Q-value $Q_0(s,a)$ for action $a$ is calculated based on the distance between $s$ and $\bar{s}_a$, normalized such that all Q-values for state $s$ sum to 1:
{\footnotesize
\[
Q_0(s,a) = \frac{\max\Big(0, 1 - \lambda \cdot \big| s - \bar{s}_a \big| \Big)}
{\sum_{a'} \max\Big(0, 1 - \lambda \cdot \big| s - \bar{s}_{a'} \big| \Big)}
\]
}
Here, $\lambda$ is a tunable linear decay coefficient that scales Q-values to decrease with distance from the target severity. Actions with target severity closer to a given state receive higher Q-values, while distant actions decay linearly toward 0. The resulting piecewise-linear, Gaussian-like profile over the action space biases the early RL policy toward selecting appropriate processor states given the observed memory access pattern, while enabling Q-learning to dynamically refine the policy according to the standard Bellman update (\S\ref{subsec:rl-solutions}).

\subsubsection{\textbf{State-to-Action Policy}}
On a PRE request, the requesting core ID and DRAM bank ID index to the corresponding CBF. ARTA (1) computes the access pattern severity from the memory addresses tracked by the CBF, (2) uses the resulting severity state ($s_{t+1}$) to index the QT, and (3) select the processor state ($a_{t+1}$) with the highest Q-value ($Q(s,a)$). ARTA computes pattern severity in a specific core as a normalized score. The score is bounded between 0 and 1, calculated as the sum of the absolute second-order difference from the memory addresses tracked by the CBF. 

\begin{figure}[t]
\centering
\includegraphics[width=1.0\linewidth]{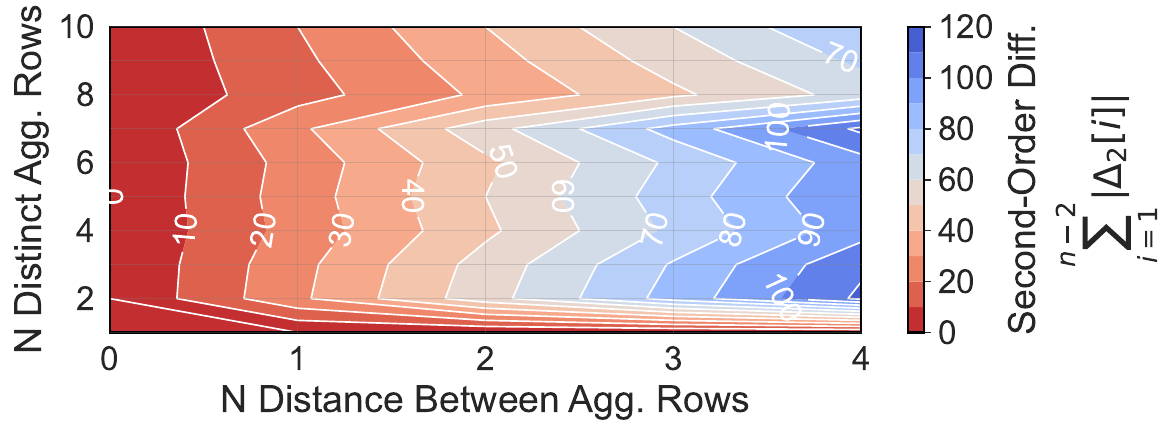}
\caption{Distribution of the sum of absolute second-order differences in a 16-entry FIFO for N-sided hammering patterns.
}
\label{fig:arta-pattern-detection}
\vspace{-4mm}
\end{figure}

Let the FIFO sequence of recent memory addresses tracked by the CBF be $F = [m_0, m_1, \dots, m_{n}]$. The first-order differences are $\Delta_1[i] = m[i] - m[i - 1], \quad 1 \le i \le n-1$ and the second-order differences are $\Delta_2[i] = \Delta_1[i] - \Delta_1[i-1], \quad 1 \le i \le n-2$.
Finally, the sum of absolute second-order differences is used to compute the normalized pattern severity score $s_{F}$:

{\footnotesize
\[
s_{F} = 1 - \min \Bigg( 1, \frac{\sum_{i=1}^{n-2} |\Delta_2[i]|}{\beta \cdot (n-2)} \Bigg), \quad s \in [0,1],
\]
}
where $\beta$ is a normalization coefficient representing the fraction of DRAM rows that an attacker can feasibly target within one $t_\text{REFW}$ window. Concretely, for a 128K-row DDR5 bank at $N_\text{BO}=64$, at most $\lfloor t_\text{REFW} / t_\text{RC} \rfloor / 64 \approx 1\%$ of rows can be hammered per window, so $\beta$ is set accordingly. This prevents benign workloads with naturally large address spreads from being incorrectly classified as attacks.
A higher score indicates a more rapid and repeated memory access pattern characteristic of RowHammer attacks. This score, referred to as the \textit{observed severity}, is adjusted to a \textit{relative severity} if the core is already throttled. The resulting score maps to a discrete severity state that ARTA uses to select the processor state with the highest Q-value at that timestep. Fig. \ref{fig:arta-pattern-detection} illustrates the distribution of the sum of absolute second-order differences for multi-sided RowHammer patterns assuming a FIFO containing up to 16 entries. The x-axis shows the distance between aggressor rows, while the y-axis denotes the N-sided pattern. The contour lines represent the sum of second-order differences, with each line indicating a set of points sharing an identical measurement. A value of 0 represents a perfect repeating pattern sequence.

Given the newly observed severity state $s_{t+1}$ and previous throttling action $a_t$, the \textit{relative severity} $s_R$ is defined as

{\footnotesize
\[
s_R =
\begin{cases}
1, & \text{if } a_t = 1, \\[6pt]
\min\!\left(1, \dfrac{s_{t+1}}{1 - a_t}\right), & \text{otherwise.}
\end{cases}
\]
}

ARTA computes the relative severity by scaling the observed severity to estimate its unthrottled equivalent. This normalization corrects for the underestimation that occurs because throttling reduces the observed severity.

At the next state transition immediately following a PRE command, ARTA updates the CBF registers. The stored severity state and throttling action are overwritten with the current state and the new action chosen by the RL policy.

\subsubsection{\textbf{Policy Evaluation}}
\begin{figure}[ht]
\centering
\includegraphics[width=1.0\linewidth]{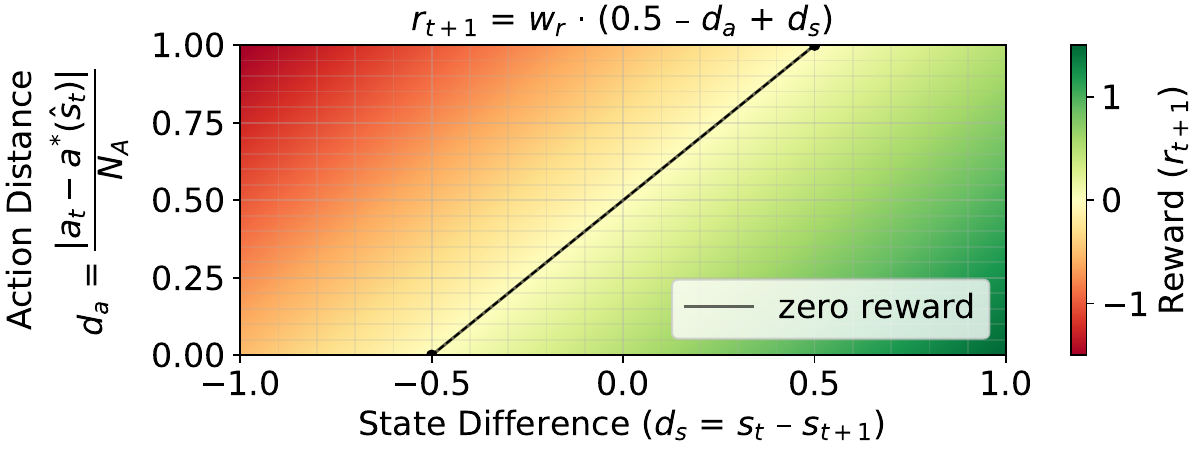}
\caption{Rewards distribution of selected actions in given states.}
\label{fig:arta-rewards}
\vspace{-5mm}
\end{figure}

Upon updating the CBF registers, ARTA updates the Q-value of the state-action pair $Q(s_t,a_t)$ corresponding to the overwritten throttling action $a_t$. This reward quantifies both the accuracy of the previous processor state and the effectiveness of mitigating high-severity memory access patterns. Integration of normalized action and state metrics simultaneously encourage the selection of processor states that (1) minimize the deviation from the optimal throttling action (i.e., highest Q-value) and (2) reduce pattern severity. As a result, the reward function generates an equally constrained range of negative and positive rewards $r_{t+1}$, as shown in Fig. \ref{fig:arta-rewards}. In the following, we describe in detail ARTA's reward function algorithm.

Let $\mathcal{S} = \{ s_0, s_1, \dots, s_{N_S} \}$ be the discrete set of severity states with cardinality $N_S$, and let $\mathcal{A} = \{ a_0, a_1, \dots, a_{N_A} \}$ be the set of available DFS operating points (P- and C-states) with cardinality $N_A$. Each continuous severity measurement is quantized by mapping the current state $s_t$ to the nearest discrete bin $\hat{s}_t \in \mathcal{S}$, where \( \hat{s}_t = \min\bigl(N_S, \max\bigl(0, \lfloor s_t \times N_S \rfloor \bigr)\bigr) \). Each action $a_i \in \mathcal{A}$ corresponds to a DFS operating point.

Given the current state $s_t$, the \textit{optimal action} $a^*(\hat{s}_t)$ is defined as
{\footnotesize
\[
a^*(\hat{s}_t) = \arg\min_{a_i \in \mathcal{A}} \, |a_i - \hat{s}_t|,
\]
}
which minimizes the absolute deviation between the target system severity and the throttling level of the chosen processor state. To reduce exploitation bias, ARTA allows selection from the top-K actions closest to the optimal action $a^*(\hat{s}_t)$ during exploration, promoting more diverse action sampling while still prioritizing reward-maximizing throttling decisions.

The \textit{normalized distance} $d_a$ between the selected action $a_t$ and the optimal action $a^*(\hat{s}_t)$ is defined as

{\footnotesize
\[
d_a = \frac{|a_t - a^*(\hat{s}_t)|}{N_A}, \quad d_a \in [0,1],
\]
}

where values near 0 indicate optimal action selection. The \textit{normalized difference} $d_s$ of the state transition is defined as

{\footnotesize
\[
d_s = s_t - s_{t+1}, \quad d_s \in [-1,1]
\]
}

where values of $0 < d_s \leq 1$ indicate an improved state. These normalized metrics quantify the accuracy ($d_a$) and effectiveness ($d_s$) of action selection $a_t$ in modifying the severity state $s_t$.


The \textit{reward} $r_{t+1}$ is computed as a weighted sum of the action and state contributions:

{\footnotesize
\[
r_{t+1} = w_r (0.5 - d_a + d_s), \quad -1.5w_r \leq r_{t+1} \leq 1.5w_r
\]
}
where $w_r > 0$ is a predefined scaling factor. Larger reward magnitudes amplify the learning signal, which can accelerate convergence by improving action differentiation. Thus, tuning $w_r$ is crucial to avoid excessive rewards that can lead to premature convergence and reduced exploration.



The reward function is incorporated into the Q-learning update algorithm as defined by the \textit{Bellman Optimality Equation} (\S\ref{subsec:rl-solutions}). The reward $r_{t+1}$ therefore serves as the immediate feedback signal that guides ARTA toward an optimal policy $\pi^*$, which maximizes cumulative expected reward over time while dynamically adjusting the clock frequency of the malicious core in response to detected RowHammer activity. The boundedness, monotonicity, and normalization of the reward ensure stable convergence of the Q-learning algorithm.

\noindent\textbf{Role of Online Q-Learning.}
The Gaussian-initialized Q-table captures a near-optimal static mapping at deployment, but online Bellman updates provide meaningful adaptation beyond this baseline. In practice, high-MPKI workloads generate address streams that naturally resemble RowHammer patterns at moderate severity; online updates allow ARTA to distinguish these benign patterns from genuine attacks over time, reducing false positives without offline retraining — the key distinction from a static lookup table.

\subsection{Technology Compatibility}

ARTA is compatible with all DDR generations by adapting $T_{A}$ to match the respective $t_\text{REFW}$ (64\,ms in DDR3/DDR4 and 32\,ms in DDR5), and is compatible with existing RowHammer mitigation mechanisms such as TRR (DDR4/DDR5) and RFM (DDR5). However, ARTA does \textit{not} strictly require integration with existing mitigation mechanisms, allowing standalone operation. To finetune ARTA's runtime operation, two tunable configuration parameters are defined: (1) the CBF size ($N_{F}$) and (2) throttling window ($T_{A}$), which govern detection accuracy and throttling rate, respectively. Increased $N_{F}$ enlarges the FIFO sampling window, generally improving detection accuracy, whereas decreased $N_{F}$ may raise false positives. Careful tuning of $N_{F}$ is required to mitigate increased system latency caused by computation of second-order differences or throttling of benign threads. Reducing $T_{A}$ enables more responsive throttling decisions, allowing the system to adapt more quickly to emerging RowHammer threats.

\begin{figure*}[htbp]
    \centering
    \includegraphics[width=1\linewidth]{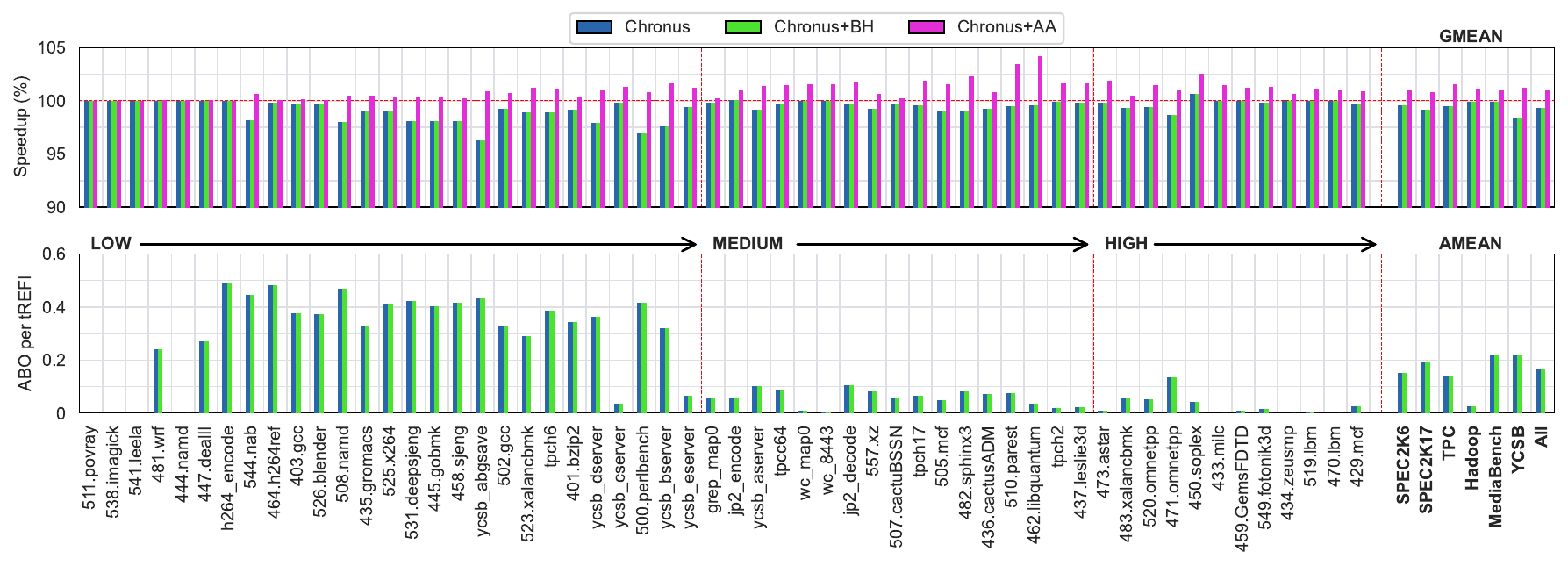}
    \vspace{-7mm}
    \caption{(a) Normalized performance (top) and (b) total ABO per $t_\text{REFI}$ (bottom) across per-trace workloads at $N_\text{BO}=64$.}
    \label{fig:perf-per-trace}
    \vspace{-5mm}
\end{figure*}

\section{Methodology}
\label{subsec:meth}
We evaluate ARTA against BreakHammer and Chronus using Ramulator2 \cite{luo2023ramulator2}, a trace-based cycle-accurate DRAM simulator (4-core x86 out-of-order CPU, 8MB shared LLC, 32GB DDR5). Table \ref{tab:system-configuration} specifies our system configuration.

\begin{table}[h]
    \vspace{-2mm}
    \scriptsize
    \centering
    \caption{Simulated System Configuration}
    \label{tab:system-configuration}
    \begin{tabular}{
        |>{\centering\arraybackslash}m{0.135\linewidth}|
        >{\raggedright\arraybackslash}m{0.765\linewidth}|
    }
        \hline
        \textbf{Processor} & 4 core, 4 GHz, 4 wide, 128 entry reorder buffer \\
        \hline
        \textbf{DRAM} & 32GB DDR5, 6400MHz DDR, 1 channel, 1 ranks, 8 bank groups, 4 banks/bank group, 128K rows/bank \\
        \hline
        \textbf{Memory \newline Controller} & 64-entry R/W queues, 8 MB LLC, 16ns $t_\text{RAS}$, 52ns $t_\text{RC}$, 410ns $\text{RFM}_\text{ab}$, $3.9\mu\text{s}$ $t_\text{REFI}$, 32ms $t_\text{REFW}$, 410ns $t_\text{RFC}$ \\
        \hline
    \end{tabular}
\end{table}

\noindent\textbf{Workloads:} Each core runs 500M instructions across 56 workloads (SPEC CPU2006/2017, TPC, Hadoop, MediaBench, and YCSB) at $N_\text{BO}$ values of 128, 64, 32, and 20. We group workloads by row-buffer misses per kilo instructions (MPKI), similar to prior works \cite{canpolat2024breakhammer, olgun2024abacus}. These groups are low [0–2), medium [2–10), and high [10+).

\noindent\textbf{Evaluation:} We compare ARTA against Chronus, a state-of-the-art DDR5 RowHammer mitigation mechanism, as well as a baseline system without mitigation, Chronus without throttling, and BreakHammer paired with Chronus. BreakHammer~\cite{canpolat2024breakhammer} reduces overheads by identifying threads that frequently trigger preventive RowHammer actions using per-thread counters updated within each 32\,ms refresh-aligned window, throttling a thread’s memory requests when its activity is disproportionately high. Chronus~\cite{canpolat2025chronus} extends the PRAC framework by enabling on-demand refresh through JEDEC-specification changes and dynamically adjusting preventive refreshes based on aggressor activity; each bank tracks its top four aggressor rows in an ATT, and Chronus refreshes the highest-priority row during each targeted RFM to minimize unnecessary refreshes while preventing RowHammer.

\noindent\textbf{Baseline Scope.} Pre-DDR5 mechanisms (PARA, Graphene, Hydra, Aqua) are excluded because they do not model the ABO/RFM timing framework of the DDR5 PRAC specification~\cite{jedecddr5c} within which ARTA operates; cross-generation comparison would require misleading assumptions. Chronus is the state-of-the-art within the DDR5 PRAC framework and is our primary comparison.

\noindent\textbf{ACPI Overhead and Overhead Inclusion.} DFS P-state transitions via Intel Speed Shift complete in $\sim$1--2\,$\mu$s; with at most one transition per 32\,ms $t_\text{REFW}$ window, amortized overhead is $<$0.007\% of execution cycles. This penalty is modeled as a fixed cycle cost in Ramulator2. All reported speedup, ABO, preventive action, RFM latency, and energy figures include ARTA’s full runtime overhead (CBF updates, Q-table lookups, severity computation, and DFS transitions).

\section{Evaluation}
ARTA's RL-based pattern detection enables precise throttling of malicious threads, effectively eliminating RowHammer attacks and enhancing system performance. In the following, we evaluate ARTA's performance based on system throughput, ABO frequency, and runtime overhead.


\subsection{Performance Improvement}


Fig. \ref{fig:perf-per-trace}a shows system performance measured by weighted speedup (\%) for workloads with MPKI in the range $[2, 10+]$ at $N_\text{BO}$ of $64$ under a multi-bank RowHammer attack from one malicious thread. Values less than $100\%$ indicate reduced system throughput. We make two observations. First, ARTA improves system throughput by a average of $1.0\%$ over baseline across all workloads with no performance degradation observed in any workload. We attribute this to ARTA’s timely suppression of RowHammer patterns that would otherwise exceed the $N_\text{BO}$ threshold, thereby avoiding the memory bandwidth loss caused by additional preventive actions. Second, ARTA outperforms BreakHammer and Chronus by a average (maximum) of $1.6\%$, improving by up to $4.2\%$ (\textit{462.libquantum}). Conversely, BreakHammer and Chronus exhibit performance degradation of up to $3.7\%$ (\textit{ycsb\_abgsave}). Our results show that the 32-sided multi-bank attack evades BreakHammer by distributing activations across all DRAM banks, thereby keeping the attacker thread’s RowHammer-preventive score close to the average score of the three benign threads. This strategy prevents the malicious thread from being identified as a statistical outlier, thus avoiding BreakHammer’s mitigation. Consequently, the attack causes no observable performance difference relative to Chronus without BreakHammer. In contrast, ARTA leverages memory address tracking for RowHammer detection, enabling reliable identification of attacks by monitoring activations at fine granularity within the row cycle time ($t_\text{RC}$), rather than the coarser refresh cycle time ($t_\text{RFC}$) granularity used by BreakHammer within throttling windows ($32ms$).

\begin{figure}[htbp]
    \vspace{0mm}
    \centering
    \includegraphics[width=1\linewidth]{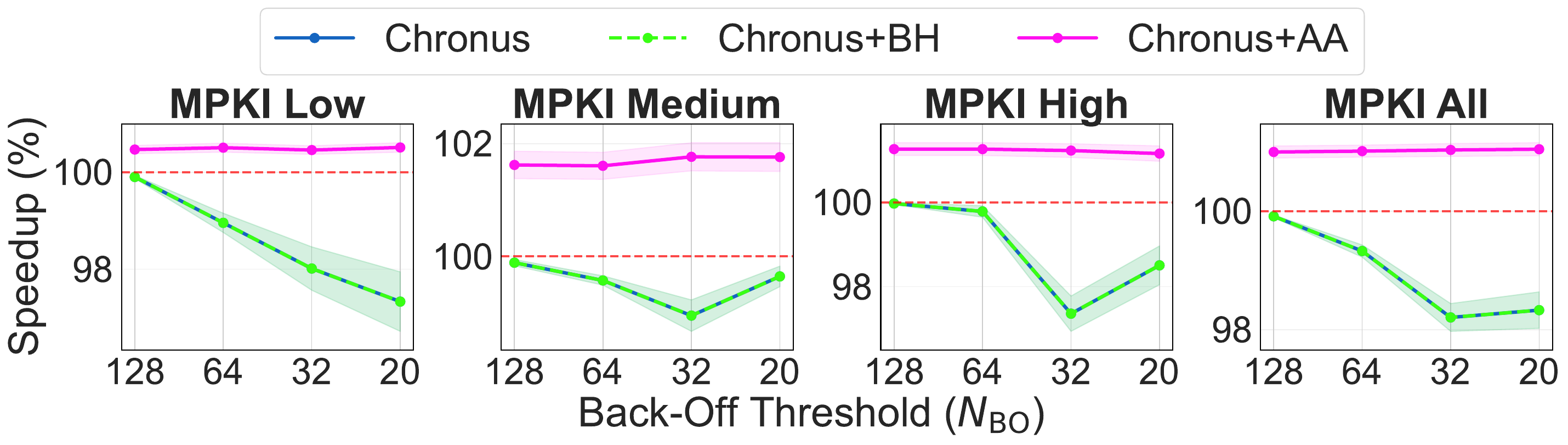}
    \caption{Per-MPKI speedup across $56$ workloads by $N_\text{BO}$.}
    \label{fig:perf}
    \vspace{-4mm}
\end{figure}

Fig. \ref{fig:perf} shows weighted speedup (\%) for each MPKI category across all evaluated $N_\text{BO}$ values from $128$ to $20$. 
We make three observations from Fig. \ref{fig:perf}. All reported performance gains are relative to BreakHammer and Chronus. First, across all workloads, ARTA achieves average (maximum)
gains of $2.7\%$ ($11.9\%$) at $N_\text{BO}$ of $20$. Second, as $N_\text{BO}$ decreases (i.e., DRAM chips more susceptible to RDE bitflips), ARTA’s maximum gains increase, ranging from $0.9\%$ at $N_\text{BO}$ of $128$ to $9.2\%$ at $N_\text{BO}$ of $32$. Third, the largest gains occur in low-MPKI workloads at $N_\text{BO}$ of $20$ ($12\%$) and $32$ ($9.2\%$), benefiting from ARTA’s pattern detection to prevent memory controller saturation. Medium- and high-MPKI workloads show smaller improvements, peaking at $N_\text{BO}$ of $32$ with gains of $5.4\%$ and $5.0\%$, respectively, as memory-bound execution limits ARTA’s frequency scaling benefits. Across all workloads at $N_\text{BO}$ of $128$, performance remains largely unaffected ($0.9\%$), confirming that DFS is selectively activated only under detected threats. Under benign-only workloads with no attacker present, ARTA incurs zero performance degradation, as DFS throttling is never triggered when memory access patterns remain below the severity threshold.

\subsection{ABO Frequency}
Fig. \ref{fig:perf-per-trace}b shows the ABO rate per $t_\text{REFI}$ at $N_\text{BO}=64$. In \textit{all} workloads, ARTA reduces ABO rates to near-zero ($<$0.001), a $5400\times$ improvement over BreakHammer and Chronus (average 16.8\%, max 49.0\%). ARTA achieves 0\% ABO in 37 of 56 workloads and reduces rates to 0\% in all high-MPKI and 15 of 17 medium-MPKI workloads, preventing bandwidth loss from preventive actions that saturate the memory controller.

\begin{figure}[htbp]
    \centering
    \includegraphics[width=1\linewidth]{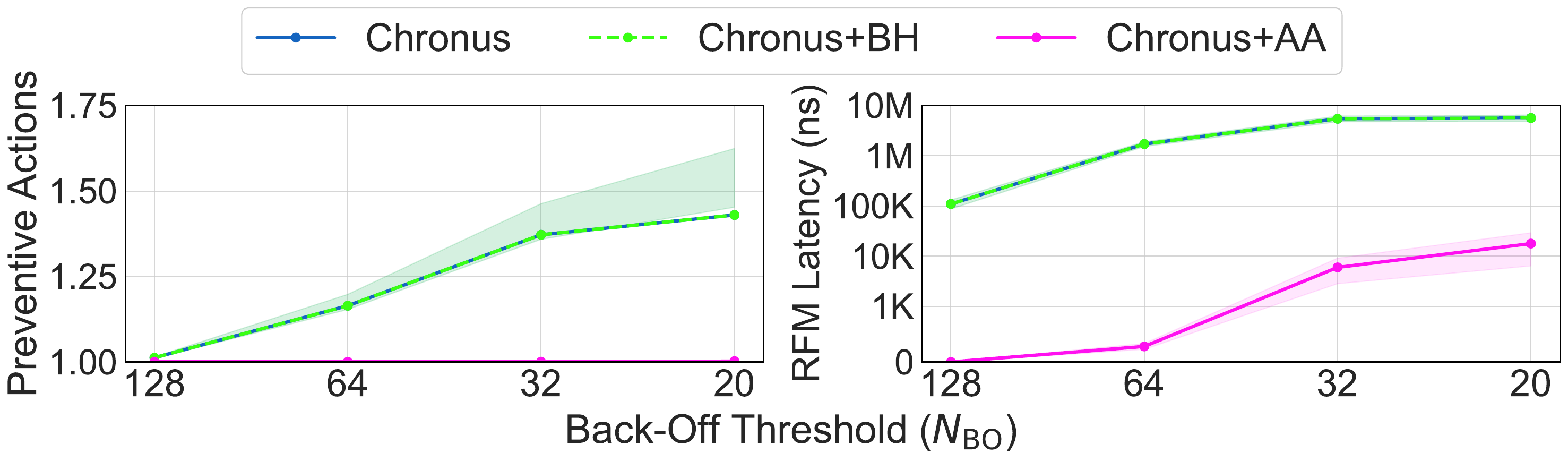}
    \caption{(a) Preventive actions (left) and (b) RFM latency (right) across $56$ workloads by $N_\text{BO}$.}
    \label{fig:prev-rfm-actions}
    \vspace{-4mm}
\end{figure}

\subsection{Performance Overhead}

Fig.~\ref{fig:prev-rfm-actions} shows preventive actions and RFM latency from $N_\text{BO}=128$ to $20$. ARTA reduces preventive actions by $17\times$ ($61\times$ max) and cumulative RFM latency by $325\times$ ($6600\times$ max) over BreakHammer and Chronus. At $N_\text{BO}=128$, ARTA achieves zero RFM latency; at $N_\text{BO}=64$, it issues at most two RFM commands in 19 of 56 workloads, versus avg $109\,\mu$s ($524\,\mu$s max) for competitors.

\begin{figure}[h]
    \vspace{1mm}
    \centering
    \includegraphics[width=1\linewidth]{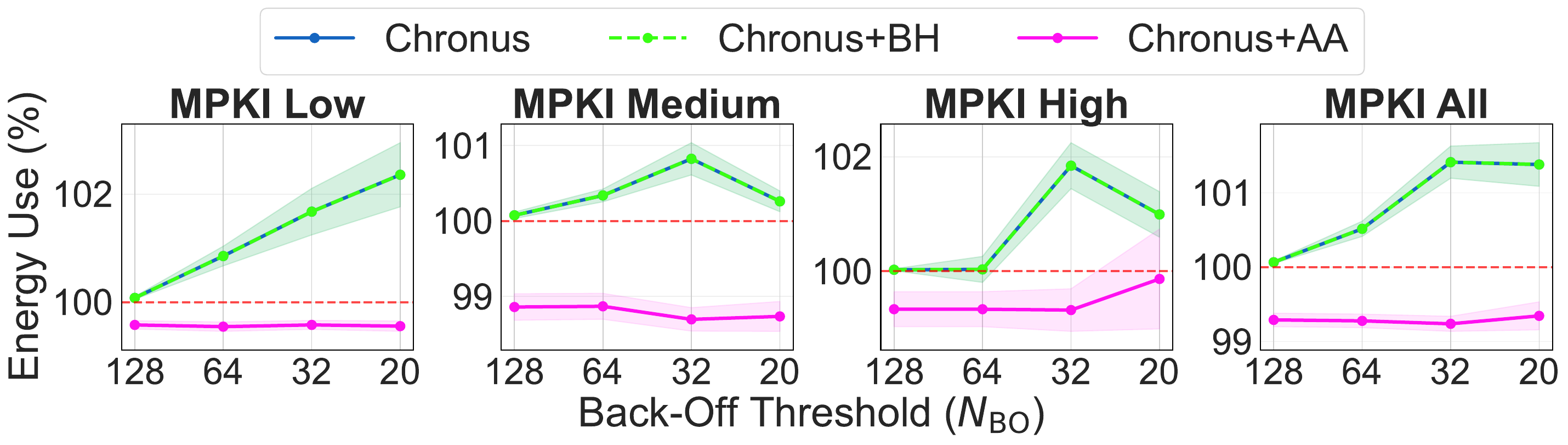}
    \caption{Per-MPKI energy consumption across $56$ workloads by $N_\text{BO}$.}
    \label{fig:energy}
    \vspace{-5mm}
\end{figure}

\subsection{Energy}
Fig.~\ref{fig:energy} reports normalized DRAM energy across MPKI categories for all $N_\text{BO}$ values. ARTA reduces energy by an average (max) of $1.43\%$ ($5.6\%$) over BreakHammer and Chronus, with peak savings in low-MPKI workloads ($1.39\%$ avg, $9.4\%$ max) and at $N_\text{BO}=32$ ($3.1\%$). By throttling malicious threads before repeated activations trigger costly RFM operations, ARTA avoids the up to $11.3\%$ energy increase incurred by BreakHammer and Chronus at low thresholds.

\begin{figure}[t]
    \centering
    \includegraphics[width=1\linewidth]{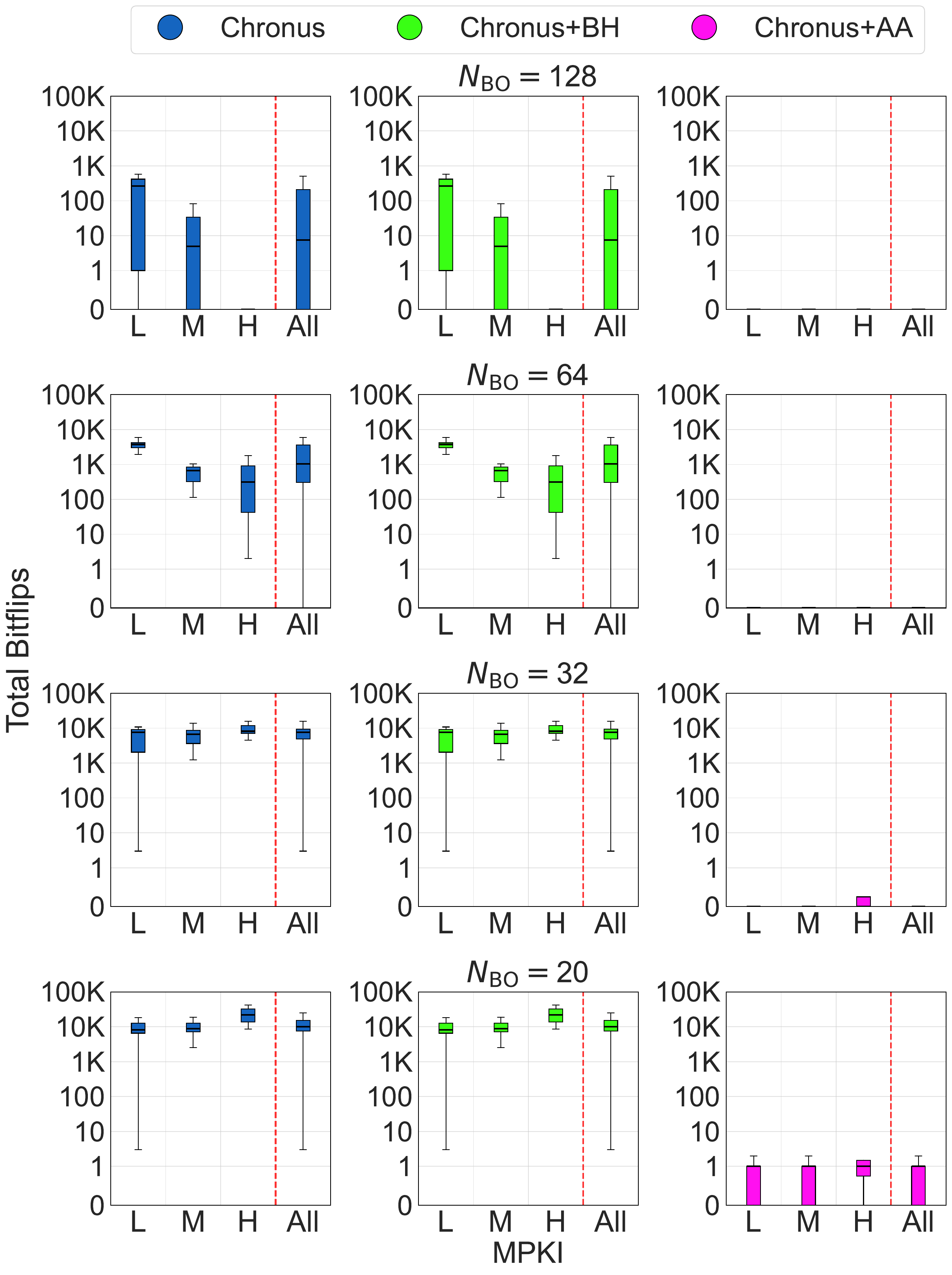}
    \vspace{-5mm}
    \caption{Total bit flips from 32-sided Multi-Bank Hammering.}
    \label{fig:total-bitflips}
    \vspace{-6mm}
\end{figure}

\section{Sensitivity Study}
\subsection{Security Evaluation}


Fig. \ref{fig:total-bitflips} shows bitflip counts for ARTA, BreakHammer, and Chronus under a multi-bank RowHammer attack across MPKI categories, shown as a box-and-whisker plot. The rows of subplots show bitflips as $N_\text{BO}$ decreases from $128$ (top) to $20$ (bottom). We make three observations from Fig. \ref{fig:total-bitflips}. First, ARTA effectively eliminates bitflips across workloads, achieving an average of $0$ for $N_\text{BO}$ values from $128$ down to $32$. In contrast, BreakHammer and Chronus exhibit bitflips at all evaluated thresholds. Second, at the most aggressive threshold ($N_\text{BO}=20$), ARTA still limits RowHammer exploitation to at most a single bitflip on average, with only $2$ bitflips observed in $6$ low- and medium-MPKI workloads. BreakHammer fails to detect and mitigate the multi-bank attack at all evaluated thresholds, producing no distinguishable differences of bitflip counts observed in Chronus. As a result, both schemes lead to system compromise, with average (maximum) bitflip counts of $8\text{K}$ ($18\text{K}$) in low-MPKI and $9\text{K}$ ($19\text{K}$) in medium-MPKI workloads. We attribute this vulnerability to the large number of preventive actions at low $N_\text{BO}$ values, which are delayed by $t_\text{REFI}$ constraints imposed by periodic refresh scheduling, allowing hammering-induced disturbances to accumulate. Third, in high-MPKI workloads, BreakHammer and Chronus exhibit the most bitflips, reaching average (maximum) counts of $22$K ($33$K) bitflips at $N_\text{BO}$ of 20 and $8$K ($12$K) at $N_\text{BO}$ of 32. ARTA, by contrast, reduces bitflips by $22\text{K}\times$ at $N_\text{BO}$ of 20, preserving data integrity under heavy memory traffic.

\subsection{Sensitivity to Configuration Parameters}
ARTA is a lightweight RL-based mitigation scheme that dynamically adjusts core throughput based on real-time memory access patterns. Unlike BreakHammer, ARTA operates fully independently by directly leveraging recent access behavior rather than monitoring preventive actions to detect and suppress RowHammer activity. ARTA uses a compact CBF to track temporal locality in memory accesses while reducing interference from inter-core and inter-bank contention.

\begin{figure}[t]
    \centering
    \includegraphics[width=1\linewidth]{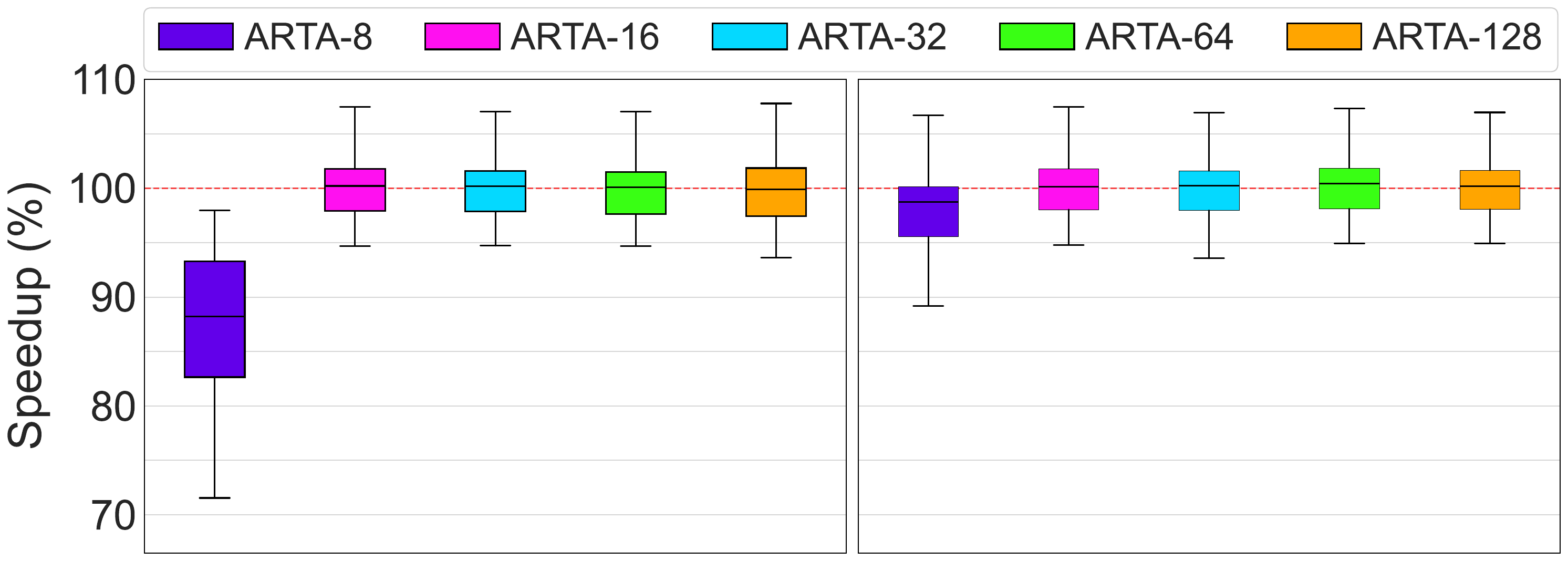}
    \caption{Speedup of ARTA (a) with PRAC at $N_\text{BO}=64$ (left) and (b) by itself (right) across $56$ multi-bank attack workloads.}
    \label{fig:fifo-sensitivity}
    \vspace{-5mm}
\end{figure}

To determine the minimum CBF size ($N_\text{F}$) that avoids false positives while maintaining high detection accuracy, we sweep $N_\text{F}$ from 128 down to 8 entries across $56$ multi-bank attack workloads at $N_\text{BO}$ of $64$. Fig.~\ref{fig:fifo-sensitivity} reports weighted speedup~\cite{eyerman2008system} for each configuration, shown as box-and-whisker subplots. We evaluate ARTA both with PRAC (left) and standalone (right) to identify the optimal configuration.


We make three observations. First, the 16-entry CBF provides the best balance of performance and security. ARTA improves performance by up to $73.6\%$ (\textit{jp2\_encode}), with benefits consistent across all MPKI categories and stable with or without PRAC. Although performance differences between CBF sizes are small, the 16-entry CBF is the most secure, maintaining $100\%$ precision and only $21$ bitflips across all workloads, compared to $37$ for the 32-entry CBF and up to $145$ for the 128-entry CBF. Larger CBFs incur additional memory accesses to update their counters, which delays throttling decisions and increases vulnerability windows, resulting in more bitflips. Second, standalone ARTA performs nearly identically to ARTA+PRAC. Performance differences remain below $0.01\%$, indicating that ARTA alone effectively throttles malicious threads, reduces DRAM contention, and preserves system throughput for benign workloads. Third, the 8-entry CBF performs the worst, with only $52.7\%$ precision. Its limited sampling capacity restricts coverage of recent memory accesses, resulting in $52$ false positives and $52\text{K}$ bitflips. High-MPKI workloads exhibit the highest false positives due to frequent row-buffer conflicts accessing different rows in the same bank, whereas low-MPKI workloads showed zero false positives. By throttling benign threads, throughput degraded by up to $87.3\%$. In all instances, both 8-entry CBF configurations performed the worst.

Based on these results, we choose $N_\text{F}=16$ because it provides the strongest security against RowHammer attacks while yielding the best overall performance, independent of underlying DRAM mitigation mechanisms. Table~\ref{tab:storage-overhead} details the per-core storage breakdown. Each 16-entry CBF stores a 17-bit row ID per entry (272\,b/bank), plus 73\,b of registers (5-bit state ID, 4-bit action ID, 64-bit clock cycle) per bank. Across 32 banks: $(272 + 73) \times 32 = 11{,}040$\,b $\approx 1.38$\,KB. The Q-Table (11 states $\times$ 6 actions $\times$ 8-bit Q-values) adds 66\,B. The total storage overhead per core is $\mathbf{1.4}$\,\textbf{KB}.

\begin{table}[h]
\centering
\caption{ARTA Storage Overhead per Core (32-bank DDR5)}
\label{tab:storage-overhead}
\scriptsize
\begin{tabular}{lrr}
\toprule
\textbf{Structure} & \textbf{Per Bank} & \textbf{Total (32 banks)} \\
\midrule
CBF FIFO (16$\times$17b row IDs) & 272\,b & 8{,}704\,b \\
Registers (5b state + 4b action + 64b cycle) & 73\,b & 2{,}336\,b \\
\midrule
CBF Subtotal & 345\,b & 11{,}040\,b (1.38\,KB) \\
Q-Table (11 states $\times$ 6 actions $\times$ 8b) & \multicolumn{2}{c}{66\,B} \\
\midrule
\textbf{Grand Total} & \multicolumn{2}{c}{\textbf{$\approx$1.4\,KB per core}} \\
\bottomrule
\end{tabular}
\vspace{-3mm}
\end{table}

\subsection{Threat Model Limitations}

ARTA has two explicit scope limitations. First, \textbf{multi-core colluding attacks}: ARTA monitors each core independently, so a coordinated attack distributing hammer activations across multiple cores — each individually below ARTA's severity threshold but collectively exceeding $N_\text{RH}$ — would not be detected. This threat class is acknowledged by BreakHammer~\cite{canpolat2024breakhammer} in its security analysis; cross-core severity aggregation at the MC is left as future work. Second, \textbf{adversarial evasion}: an attacker aware of the second-order difference metric could inject dummy accesses to inflate the difference sum and suppress the severity score. ARTA's relative severity normalization partially compensates, but formal guarantees against a fully adaptive attacker require future evaluation. A quantitative \textbf{fairness} analysis of per-thread throughput unfairness under multiprogram workloads is also deferred to future work.

\section{Related Work}

Throttling defenses limit access rates before $N_\text{RH}$: BreakHammer~\cite{canpolat2024breakhammer} tracks per-thread preventive actions but misses coordinated multi-bank attacks~\cite{kang2024sledgehammer}; BlockHammer~\cite{yauglikcci2021blockhammer} operates independently of refresh mechanisms. Deterministic counter-based schemes~\cite{park2020graphene,qureshi2022hydra,bostanci2024comet,olgun2024abacus,qureshi2025moat,woo2025qprac} refresh rows precisely but suffer blind spots or excessive all-bank refreshes at low thresholds; Chronus~\cite{canpolat2025chronus} refines PRAC timing yet remains exploitable. Probabilistic approaches~\cite{qureshi2024mint,jaleel2024pride} reduce storage but incur ${\sim}30\%$ bandwidth loss at $N_\text{BO}=250$~\cite{woo2025qprac}. Software defenses~\cite{zhang2022softtrr,van2018guardion,aweke2016anvil,bock2019rip} lack DRAM parameters; ECC~\cite{fakhrzadehgan2022safeguard,kim2023kill} cannot correct multi-bit flips. Prior ML detection~\cite{joardar2022machine} requires offline supervised training. ARTA is the first RL-based throttling mechanism: it learns policies online without DRAM-side changes, detects multi-bank attacks missed by heuristic defenses, and reduces refresh demand by suppressing activations before $N_\text{BO}$.

\section{Conclusion}
ARTA is the first RL-based throttling mechanism for RowHammer mitigation. It adds only 1.4\,KB of SRAM per core to the memory controller, learns online without DRAM-side modification, and detects multi-bank attacks that evade existing defenses — eliminating nearly all bitflips down to $N_\text{BO}=64$, reducing RFM latency by $325\times$, and improving performance by up to $73.6\%$ over Chronus and BreakHammer. Cross-core colluding attacks and adversarial evasion via dummy accesses remain open threat model gaps addressed in future work.

\bibliographystyle{IEEEtran}
\bibliography{references}

\end{document}